\documentclass{moriond}

\def\be{\begin{equation}}
\def\ee{\end{equation}}
\def\bea{\begin{eqnarray}}
\def\eea{\end{eqnarray}}

\newcommand{\Photo}{}

\begin{document}
\vspace*{4cm}
\title{PROBES FOR CP VIOLATION IN B DECAYS AT THE FCC: \\
A THEORIST'S PERSPECTIVE}

\author{R. FLEISCHER}

\address{Nikhef, Science Park 105, 1098 XG Amsterdam, Netherlands and 
Department of Physics and Astronomy, Vrije Universiteit Amsterdam, 1081 HV Amsterdam, Netherlands}

\maketitle\abstracts{CP violation offers powerful probes to explore the quark-flavour sector, where decays of $B$ mesons have been key players since decades. I discuss a variety of probes ranging from non-leptonic to rare $B$ decays, offering exciting opportunities at the FCC 
in the era after the HL-LHC and Belle II.}

\section{Setting the Stage}
The quark-flavour sector of the Standard Model (SM), allowing us to accommodate CP violation through the Cabibbo--Kobayashi--Maskawa (CKM) matrix, shows a very rich phenomenology.\cite{SF} The Baryon asymmetry of the Universe indicates that we are missing CP violation, while extensions of the SM usually contain also new sources of CP violation. Studies of CP violation in the quark-flavour sector probe genuinely very high energy scales of New Physics (NP). The limitation of this tool to search for indirect signals of physics beyond the SM is given by precision, in contrast to the direct NP searches which are limited by the centre-of-mass energy of the used particle collider.

For decades, key players in the exploration of CP violation have been decays of $B$ mesons, which have led to a lot of theoretical work and strategies.\cite{RF-CP-rev} The experimental exploration is currently pursued in a very impressive way by LHCb as well as ATLAS and CMS at CERN's LHC and Belle II at the Super KEK B factory, providing a wealth of experimental results. Exciting years are ahead of us while moving forward along the high-precision frontier thanks to the full exploitation of Belle II and the HL-LHC with the LHCb upgrade II. Interesting opportunities for heavy-quark physics arise also at the FCC in the era after the LHC.\cite{MW-21}

Questions of the following kind come to mind concerning the status of particle physics around the year 2050: Will discrepancies with respect to the SM have been established? Will new particles have been discovered? Will new sources of CP violation have been revealed? What about the theoretical status and progress? I'm of course not a fortune teller and time will tell. Future scenarios were discussed as input to the 2026 update of the European Strategy for Particle Physics.\cite{Physics-Briefing-Book} Illustrations of the spectacular expected progress for the determination of the Unitarity Triangle (UT) of the CKM matrix were also given with projections up to the FCC-ee around 2050. Interestingly, key limitations are given by the CKM matrix elements $|V_{ub}|$ and $|V_{cb}|$, where we are unfortunately facing long-standing tensions between inclusive and exclusive determinations.\cite{UT-NP}  The FCC-ee will also have a key impact on this topic. 

CP violation in $B$ decays provides a central input to determine the apex of the UT but also beyond in view of testing the SM and detecting possible new sources of CP violation. In the following, I present a selection of $B$ decays which appear very interesting for the FCC-ee.

\boldmath
\section{Decays of the kind $B^0_{d}\to J/\psi K_{\rm S}$, $B^0_s\to J/\psi \phi$}\label{sec:BJpsiK}
\unboldmath
Pushing the high-precision frontier forward, it is crucial for resolving smallish effects from physics beyond the SM to have a critical look at theoretical analyses and their approximations. In the case of decays of $B$ mesons, the key issue is related to the impact of strong interactions. The goal and challenge is to match the experimental and theoretical precisions. Concerning the ``golden" decays $B^0_d\to J/\psi K_{\rm S}$ and $B_s^0\to J/\psi \phi$, which allow the determination of the $B^0_{d,s}$--$\bar B^0_{d,s}$ mixing phases $\phi_{d,s}$, the experimental prospects for the precision up to the FCC-ee are very impressive, as highlighted in the
Physics Briefing Book.\cite{Physics-Briefing-Book} Theoretical uncertainties arise from doubly Cabibbo-suppressed penguin contributions.\cite{RF-97,RF-99} The experimental precision requires the control of these effects to reveal possible NP contributions. This topic has received long-standing interest in the community.\cite{RF-CP-rev} Since the hadronic corrections to the experimental values of $\phi_{d,s}$ cannot be reliably calculated within QCD, we employ  control channels to take them into account. The key example is the $B^0_s\to J/\psi K_{\rm S}$ decay, which is related to $B^0_d\to J/\psi K_{\rm S}$ through the $U$-spin symmetry of strong interactions.\cite{RF-99} Further control channels as well as the $B^0_s\to J/\psi \phi$ decay can be added to the analysis, resulting in a complex strategy to take the penguin effects in the determinations of $\phi_d$ and $\phi_s$ into account.\cite{pen-25} Assumptions made to simplify the analysis and lack of data leave a lot of room for future improvements. Decays of the kind $B^0_{d}\to J/\psi K_{\rm S}$ and $B^0_s\to J/\psi \phi$ should be fully exploited at the FCC-ee. The control of the hadronic penguin effects will be crucial and equally important to improving precision on the golden modes, as is the case for the full exploitation of the HL-LHC and Belle-II. Will we already reveal a SM discrepancy for $\phi_{s(d)}$ in the next years? Which precisions could be achieved at the FCC-ee?

\boldmath
\section{Decays of the kind $B^0_{d}\to \pi^0 K_{\rm S}$}
\unboldmath
Decays of $B$ mesons into $\pi K$ final states are governed by QCD penguin topologies and also electroweak (EW) penguins may play a significant role. Sophisticated strategies were developed utilising the whole $B\to\pi K$ system as well as $B \to \pi\pi$ modes, 
as discussed in more detail in the contribution by Eleftheria Malami.\cite{EM-Proc} The decay $B^0_{d}\to \pi^0 K_{\rm S}$ is a particularly interesting probe for testing the SM: Utilising an isospin relation between the neutral $B\to\pi K$ amplitudes with a minimal further $SU(3)$ input, SM correlations between the direct and mixing-induced CP-violating observables of $B^0_{d}\to \pi^0 K_{\rm S}$ can be calculated.\cite{FJPZ} Interestingly, the experimental data result in a puzzling situation.\cite{EM-Proc,FJPZ,FJMV-18}  Will we already manage to establish SM discrepancies in the next years through Belle II? EW penguin topologies offer an exciting portal for NP effects to enter the $B^0_{d}\to \pi^0 K_{\rm S}$ channel,\cite{FJPZ,FJMV-18} where models with extra $Z'$ bosons offer attractive scenarios.\cite{BA} At the FCC, the interplay with rare $B$ decays appears particularly interesting, and the question arises whether this new collider will lead to the observation of $Z'$ bosons (and other associated new particles). Examples of further interesting modes governed by penguin topologies are $B^0_d\to \phi K_{\rm S}$ and $B^\pm\to \phi K^\pm$ decays.\cite{FM,FGV}  Which precisions could be achieved at the FCC-ee?

\boldmath
\section{Decays of the kind $B^0_s\to D_s^\mp K^\pm$, $B^0_d\to D^\mp\pi^\pm$}
\unboldmath
Utilising CP violation in the pure tree decays $B^0_s\to D_s^\mp K^\pm$ and $B^0_d\to D^\mp\pi^\pm$, we may determine the CP-violating phases $\phi_s+\gamma$ and $\phi_d+\gamma$, respectively, in an essentially theoretically clean way.\cite{ADK,RF-BsDsK} The UT angle $\gamma$ can be extracted with the help of $\phi_{d,s}$ (see Section~\ref{sec:BJpsiK}). Interestingly, data for branching ratios and CP violation in the $B^0_s\to D_s^\mp K^\pm$ system result in a puzzling situation, which is also present in decay rates of other $B$ decays with similar dynamics.\cite{RF-CP-rev,FM-BsDsK-1,FM-BsDsK-2} Should this situation be related to NP effects, they would have to arise at the decay amplitude level. It will be very interesting to monitor how the situation will evolve at the LHCb Upgrade II. These decays offer amazing probes for testing the SM and searching for new sources of CP violation.  The theoretical precision of the extraction of $\gamma$ is limited by second-order electroweak corrections, which result in an uncertainty below the $10^{-7}$ level.\cite{BZ} Concerning the search for possible new sources of CP violation, it will be crucial to individually measure $\gamma$ through time-dependent measurements such as $B^0_s\to D_s^\mp K^\pm$  and time-independent $B\to DK$ methods with highest accuracies. Will tensions arise guiding us to NP? In this endeavour, also $B^0_d\to D^\mp\pi^\pm$ decays may offer further valuable insights. 
Which precisions could be achieved at the FCC-ee? For a study of CP violation in $B^0_s\to D_s^\mp K^\pm$ 
decays at the CEPC, see Ref.~\cite{CEPC}.

\boldmath
\section{Decays of the kind $B^0_{s(d)}\to \ell^+\ell^-$, $B\to K^{(*)}\ell^+\ell^-$}
\unboldmath
Rare $B$ decays provide yet another most interesting tool with new opportunities to explore CP violation at the FCC-ee. In the SM, these processes originate from penguin and box topologies and are strongly suppressed. The leptonic $B^0_{s(d)}\to \ell^+\ell^-$ channels belong to the theoretically cleanest decays. They have high sensitivity to physics from beyond the SM. In view of their helicity suppression, they are particularly powerful probes for new (pseudo)-scalars. So far only the $B^0_{s}\to \mu^+\mu^-$ decay has been observed at the LHC, with first limits on $B^0_{s}\to \tau^+\tau^-$.  It is important to search also for $B^0_{s(d)}\to e^+e^-$ modes, which are enormously helicity suppressed in the SM but could be enhanced through NP effects; their observation would be a spectacular signal of beyond the SM physics.\cite{Bsee-paper}  The neutral $B^0_{s(d)}\to\ell^+\ell^-$ decays show intriguing time-dependent interference effects 
arising from $B^0_{s(d)}$--$\bar B^0_{s(d)}$ 
 mixing,\cite{Bsmumu-ADG} in analogy to phenomena in non-leptonic $B$ decays.\cite{RF-CP-rev} 
 Concerning the $B^0_s\to\mu^+\mu^-$ mode, 
 the sizeable decay width difference $\Delta\Gamma_s$ of the neutral $B_s$ system provides access to an untagged observable 
 $\mathcal{A}_{\Delta\Gamma_s}$, playing an important role to complement  the constraints on NP from the CP-averaged branching 
 ratio.\cite{Bsmumu-ADG} The measured $B^0_s\to\mu^+\mu^-$ rate is consistent with the SM within the uncertainties but still leaves a lot of room for possible new effects in the (pseudo)-scalar sector, including CP violation. It will be important to measure  $\mathcal{A}_{\Delta\Gamma_s}$ with highest precision,\cite{Bsmumu-CPV} following the first pioneering measurements at the LHC, to reveal possible NP contributions. Further exciting opportunities arise through time-dependent CP asymmetries,\cite{Bsmumu-CPV,CPV-rare-B} which require tagging information. 
 
 The studies of leptonic $B^0_s\to\mu^+\mu^-$ decays are complemented in a synergetic way through analyses of CP violation in 
 $B^0_d\to K_{\rm S}\mu^+\mu^-$, $B\to K^*\mu^+\mu^-$, $B^0_s\to \phi \mu^+\mu^-$ modes. In particular, a short-distance coefficient 
 $ \mathcal{C}_{10}^- $ is required as input, which can be extracted through observables provided by semi-leptonic rare $B$ decays and 
 CP asymmetries.\cite{RF-BVellell} Interestingly, CP violation in $B\to K^{(*)}\mu^+\mu^-, K^{(*)} e^+e^-$ decays offers also new perspectives for testing electron--muon universality, where current data with $R_{K^{(*)}}\sim1$ still leave significant space.\cite{RK-CPV} The first measurement of time-dependent CP violation in $B^0_d\to K_{\rm S}\mu^+\mu^-$ by LHCb is a very important first step.\cite{LHCb-CP-rare} Explorations of CP violation in rare $B$ decays are a complex topic and offer a very interesting playground for future experiments. Which precisions could be achieved at the FCC-ee?

\section{Concluding Remarks}
CP violation in $B$ decays provides fantastic opportunities, where the processes discussed above are just a selection. The FCC-ee should aim to fully exploit these NP probes, confirm results from the HL-LHC and Belle II, and enter new unknown territory around 2050, which will also provide guidance for the FCC-hh. Detailed feasibility studies and benchmark scenarios are needed and actually explored in the current Flavours at FCC Workshop at CERN.\cite{FlavourFCC-Workshop} It will be interesting to look back at the processes and observables discussed in this presentation in 2050.

\section*{Acknowledgments}
I am very grateful to my students and collaborators for the enthusiasm and inspiring work on our projects. Research presented in this proceedings contribution was supported by the Netherlands Organisation for Scientific Research (NWO). I would like to thank Nazila Mahmoudi and her co-organisers for the invitation and organising the wonderful conference.

\section*{References}

\end{document}